\documentclass[twocolumn,apl,amsmath,amssymb,showpacs]{revtex4-2}
\usepackage{epsf}
\usepackage[utf8]{inputenc}
\usepackage{amsmath}
\usepackage{amsfonts}
\usepackage{amssymb}
\usepackage{makeidx}
\usepackage{color}
\usepackage{gensymb}
\usepackage{graphicx}
\usepackage{bm}
\usepackage{soul}
\usepackage{sidecap}
\usepackage{mathtools}
\usepackage{geometry}
\usepackage{multirow}
\usepackage{array}
\usepackage{outlines}
\begin{document}

\title{\normalsize Electrochemical investigation of MoSeTe as an anode for sodium-ion batteries}

\author{Priya Mudgal} 
\affiliation{Department of Physics, Indian Institute of Technology Delhi, Hauz Khas, New Delhi-110016, India}
\author{Himani Arora} 
\affiliation{Department of Physics, Indian Institute of Technology Delhi, Hauz Khas, New Delhi-110016, India}
\author{Jayashree Pati}
\affiliation{Department of Physics, Indian Institute of Technology Delhi, Hauz Khas, New Delhi-110016, India}
\author{Manish K. Singh}
\affiliation{Department of Physics, Indian Institute of Technology Delhi, Hauz Khas, New Delhi-110016, India}
\author{Mahantesh Khetri}
\affiliation{Department of Physics, Indian Institute of Technology Delhi, Hauz Khas, New Delhi-110016, India}
\author{Rajendra S. Dhaka}
\email{rsdhaka@physics.iitd.ac.in}
\affiliation{Department of Physics, Indian Institute of Technology Delhi, Hauz Khas, New Delhi-110016, India}

\date{\today}      

\begin{abstract}

Sodium ion batteries (SIBs) are considered as an efficient alternative for lithium-ion batteries (LIBs) owing to the natural abundance and low cost of sodium than lithium. In this context, the anode materials play a vital role in rechargeable batteries to acquire high energy and power density. In order to demonstrate transition metal dichalcogenide (TMD) as potential anode materials, we have synthesized MoSeTe sample by conventional flux method, and the structure and morphology are characterized using x-ray diffraction (XRD), field-emission scanning electron microscopy (FESEM), transmission electron microscopy (TEM), and Raman spectroscopy. These characterisations confirm the hexagonal crystal symmetry with p63/mmc space group and layered morphology of MoSeTe. We investigate the electrochemical performance of a MoSeTe as a negative electrode (anode) for SIBs in the working potential range of 0.01 to 3.0~V. In a half-cell configuration, the MoSeTe as an anode and Na metal as counter/reference electrode exhibits significant initial specific discharge capacities of around 475 and 355 mAhg$^{-1}$ at current densities of 50 and 100 mAg$^{-1}$, respectively. However, the capacity degraded significantly like $\approx$200~mAhg$^{-1}$ in 2nd cycle, but having $\approx$100\% Coulombic  efficiency, which suggest for further modification in this material to improve its stability. The cyclic voltammetry (CV) study reveals the reversibility of the material after 1st cycle, resulting no change in the initial peak positions. The electrochemical impedance spectroscopy (EIS) measurements affirms the  smaller charge transfer resistance of fresh cells than the cells after 10th cycle. Moreover, the extracted diffusion coefficient is found to be of the order of 10$^{-14}$ cm$^2$s$^{-1}$.

\end{abstract}


\maketitle

\section*{\noindent ~Introduction}

The meteoric rise of eco-friendly renewable energy sources has come up with notable extensions in the application of biomass, hydro, solar, and wind resources. However, these are only be used if the climate, location, etc. are suitable \cite{earth}. The evolution of advanced conversion technologies and high-level energy storage is required to balance the increasing demands for electric vehicles, transportable electronic equipment, etc. In this line, rechargeable batteries play an important role in achieving renewable energy rather than fossil-based fuels have largely progressed in the area of providing energy and giving opportunities for scientists to perform applications globally \cite{r23, intro2}. The lithium-ion batteries (LIB) are considered best for energy storage applications; however, due to the lack of lithium resources and nonuniform distribution, there is need to search for alternative \cite{i3, i4}. In recent years, sodium-ion batteries (SIB) are considered to be a better alternative to LIBs, owing to the uniform natural availability of sodium on the earth, which significantly reduces the cost of SIBs with a minimum compromise on performance \cite{earth, r19}. The SIBs possess the similar electrochemical mechanism of intercalation/de-intercalation as that of LIBs \cite{i5, r20}, and therefore, they evoked among the research community to achieve the milestone in terms of energy density and power density. In the case of LIBs, graphite is used as anode material, but cannot be used for SIBs as it is not thermo-dynamically stable with sodium \cite{Hwang2017, r21}. The challenges with SIBs are because of large ionic radii of Na$^{+}$ (1.02 \AA)~ as compared to Li$^{+}$ (0.76 \AA) and that is why many other electrode materials are not compatible with SIBs \cite{r19, r22}. Also, owing to the large ionic radius of Na$^{+}$, stability and capacity of SIBs is lower than its lithium equivalent. This reduces the diffusion of Na$^{+}$ and offers large volume change during the mechanism of sodiation/desodiation, and hence the degradation of the structure of the host material limits their application in the energy storage devices \cite{i6}. So far various anode materials have been examined to check their suitability for SIBs. Therefore, finding a layered anode material is crucial to abolish the limitations of graphite in SIBs for energy storage applications \cite{intro1, i7, i8, i9, PatraAEM21}.  

In today’s era, transition metal dichalcogenides (TMDs), particularly MoTe$_2$ and MoSe$_2$, are found to be a very effective material for energy storage devices owing to their large surface area and layered structure \cite{i14,i15,i16,i17}. The TMDs have a general structure of type MX$_2$, where M is a metal squeezed in the middle of X atoms making an X-M-X layer. The bond M-X is a covalent bond where X belongs to the oxygen family, also known as chalcogens. The conventional TMDs like MoSe$_2$, MoTe$_2$, WS$_2$ acquire larger inter-planar distance of around 0.65~nm \cite{i10}. In general, TMDs provide easy intercalation/de-intercalation of sodium ions owing to effortless interlayer expansions, but limit cyclic stability as volume changes largely \cite{i13}. In addition, the TMDs show high theoretical capacity and high conductivity, which are important factor to decide battery performance. These materials undergoes various types of reactions subject to provide the voltage windows; for example, conversion and insertion reactions. Also, the synthesis methods like hydrothermal/solvo-thermal/flux are crucial to control the structure and the morphology to improve the electrochemical performance. However, the flux method is having many merits as no chemical solution or no flow of toxic gases (selenium) are required, and we get clean shiny sample in this process, discussed below. 

Further, in the case of Janus TMDs, one X is replaced by another atom of the same group. The two adjacent layers of the X-M-X or X-M-Y are linked one another with weak Vander waals force, and therefore the host material acquires higher interstitial sites \cite{i11}. Interestingly, the bilayer Janus materials are good for energy storage applications because volume change is small during the sodiation/de-sodiation process that results in high storage capacity. Further, anode conductivity can be enhanced due to the transition of janus monolayers from semiconducting nature to metallic nature during the adsorption of the sodium ion \cite{i18}. To the best of our knowledge, MoSeTe related electrochemical studies are not reported for SIBs till date. Therefore, in this paper, we investigate the electrochemical performance of MoSeTe anode, which is a Janus TMD and prepared by conventional flux method. It exhibits significantly higher specific discharge capacity of around 200 mAhg$ ^{-1} $ in the second cycle. The charge transfer resistance of fresh cells is found to be around 30-40~\ohm, which increases with number of cycles.  Moreover, we extracted the diffusion coefficient which is of the order of 10$^{-14}$ cm$^2$s$^{-1}$.

 \section*{Experimental Details:}

\textbf{Synthesis of MoSeTe using flux method}: 

For the synthesis of MoSeTe sample through flux growth \cite{e1}, all precursors were taken in stoichiometric proportion of Se (99.999\%) and Te (99.995\%) pieces, and Mo foil (99.99\%) from Alfa Aesar or Sigma Aldrich. These precursors were mixed in the ratio of 1:1:1 (0.9515~g, 0.7831~g, and 1.2654~g) and sealed in a quartz tube ampoule under rough vacuum. We employed double sealing quartz tubes to avoid breakage during quenching. This double-sealed quartz tube was put into a furnace (from Nabertherm GmbH, Germany) and the temperature was increased up to 1100$\degree$C with the rate of 60$\degree$C/h. The sample was kept at 1100$\degree$C for 12 hrs, and then cooled down to 900$\degree$C at a rate of 1$\degree$C/hr, following the optimized heating profile \cite{e1}. After that, the sample was quenched into cold water. We obtained the shiny pieces after quenching the sample, which were used to make powder and store in the vacuum desiccator for further use. 

\textbf{Materials characterisation:}
 
 The x-ray diffraction (XRD) measurements were carried out at room temperature with Cu-K$\alpha$ radiations (1.5406 \AA) using Panalytical Xpert3 X-ray diffractometer in the range of 10$\degree$ to 90$\degree$. The surface morphology of the prepared powdered sample was investigated by FE-SEM using the apparatus from JEOL JSM-7800F prime system. The HR-TEM measurements were performed at Tecnai G2-20 system and the data were analysed through Image-J software. The Raman spectra were recorded using a confocal microscope of 532~nm Ar-laser with a power of 1~mW and 2400 lines/mm grating. 

\textbf{Coin cell fabrication:}

We have fabricated coin cells with 2032-type to investigate the suitability of the material MoSeTe as an anode for sodium-ion batteries. The slurry was prepared by taking 70$\%$ weight of as prepared MoSeTe as active material, 15$\%$ weight of the carbon black to increase the conductance, 15$\%$ weight of PVDF (polyvinylidene difluoride) binder and a solvent named N-Methyl Pyrrolidine (NMP). After this process, the prepared solution was stirred for 12 hr to obtain a uniform mixture, which was then coated on Cu foil (current collector). The coated sheet was vacuum dried at 110$\degree$C for 12 hrs. A disc cutter was used to cut the electrodes from the Cu coated sheet, having a diameter of 15 mm. Here, the mass loading of active material was 2~mg/cm$^{2}$. As prepared electrodes were vacuum dried at 80$\degree$C for 8 hrs. We assembled three cells, of approximately similar weight of active material, inside the glovebox (UniLab Pro SP from MBraun, Germany) with O$_2$ and H$_2$O content less than 0.1 ppm. In the fabrication process, we have used 1M NaClO$_4$ with EC and DEC in 1:1 ratio as electrolyte, glass fiber (GB100R) as separator and Na foil as reference/counter electrode. The amount of electrolyte used in the fabrication of the half-cell was nearly 0.2 ml.  

\textbf{Electrochemical Testing:}

The electrochemical impedance spectroscopy (EIS) data were collected in a frequency range of 100~kHz to 10~mHz with an AC voltage amplitude of 10~mV using VMP-3 potentiostat from Biologic. The cyclic voltammetry (CV) measurements were performed in a potential window of 0.01V to 3V at a scan rate of 0.05 mVs$^{-1}$. The galvanostatic charge-discharge measurements were carried out through Neware battery analyzer BTS400 at different current densities. 

\section*{Results and Discussion }

\begin{figure*}
    \centering
\includegraphics[width=14cm,height= 10cm]{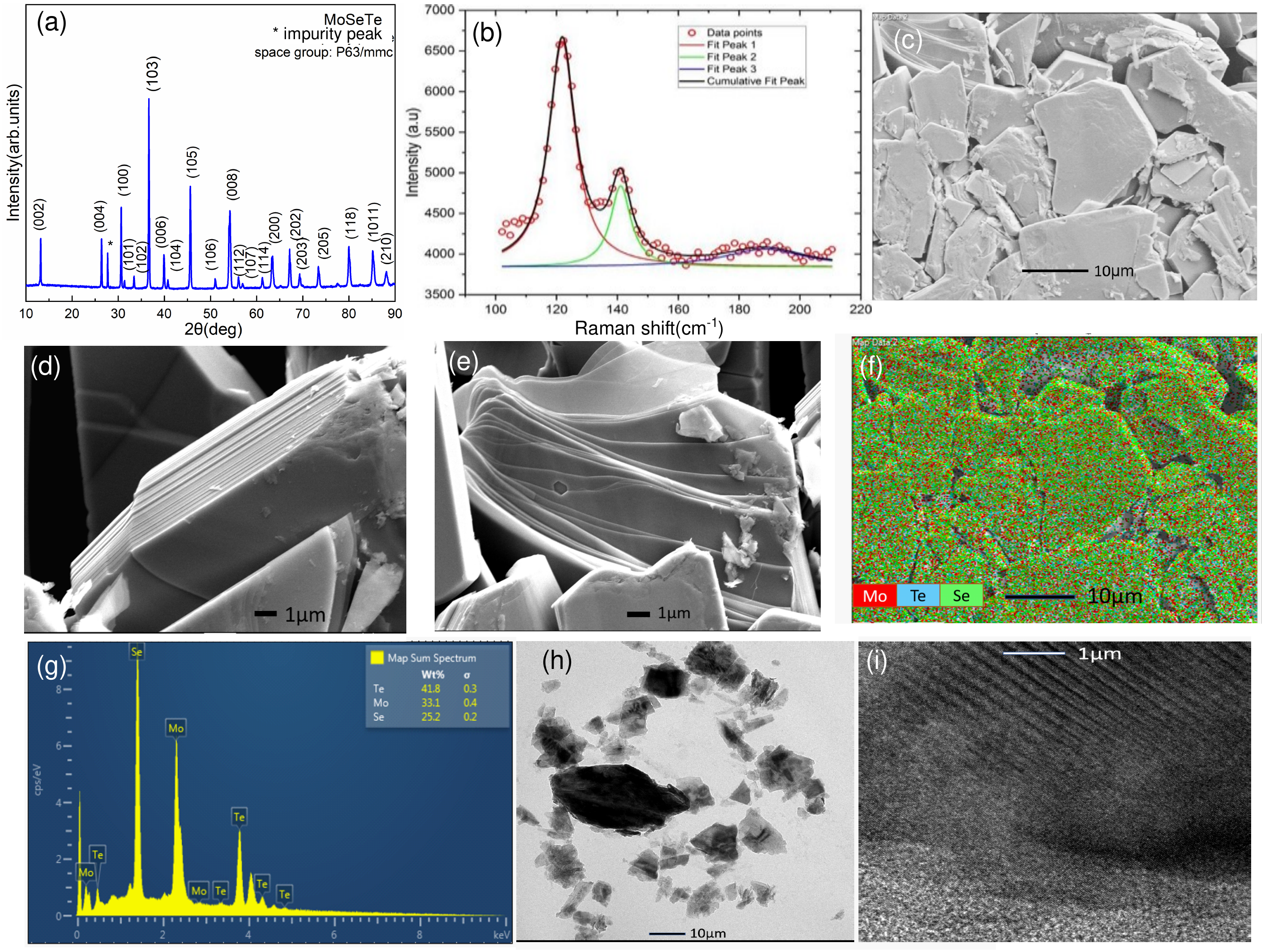}
	\renewcommand{\figurename}{Figure}
	\caption {The structural and morphological characterisations of as-prepared MoSeTe: (a) the XRD pattern with indexed planes, (b) the Raman spectrum (red circles), fitting (black line) and de-convoluted peaks, (c-e) the FE-SEM micrographs of the grown MoSeTe crystal revealing layered structure,  (f, g) the EDX of the corresponding image of FE-SEM, (h) the TEM image, and (i) the magnified TEM image showing fringes.}
	\label{ch}
\end{figure*}

From the XRD study in Figure~\ref{ch}(a), the Bragg diffraction pattern of MoSeTe is found to be well-matched with the MoTe$_2$ XRD pattern \cite{e1}. By comparing critically with the patterns reported for MoTe$_2$ and MoSe$_2$, it is evident that the sharp peaks in the XRD pattern of MoSeTe indicate high crystallinity and confirm the space group P63/mmc having 2~H phase \cite{e1}. The presence of (002), (004), (100), (103) diffraction peaks in Figure~\ref{ch}(a) can be indexed to the 2~H MoTe$_2$ phase \cite{ROY_ACS_16, ZANG_NANOSCALE_17, KWON_ACS_21}. A small additional peak marked by star in Figure~\ref{ch}(a) can be due to some impurity of unreacted part. The crystal structure of the grown MoSeTe sample was found to be hexagonal symmetry with lattice parameters $a=b=$ 3.53 \AA~ and $c=$ 13.882 \AA. The Raman spectrum in Figure~\ref{ch}(b) was recorded at room temperature and peaks were fitted using the Lorentzian peak shape function. Two Raman active modes are found to be at 122 and 182~cm$^{-1}$ coresponding to E$_{1g}$ and A$_{1g}$, respectively, which confirm the 2-H phase of the grown sample \cite{r2}. The peak at 141 cm$^{-1}$ might be due to the activation of second-order Raman process \cite{GuoPRB15}. In Figure~\ref{ch}(c, d, e), the FE-SEM images predict the layered structure of the material, which is suitable for ease migration of Na-ion and diffusion of electrolyte. Figure~\ref{ch}(d, e) demonstrate the stacking of several layers in different orientations. The sample was quenched at 900$\degree$C and owing to its incomplete solidification, spongy-like isolated structures are obtained. The quantitative EDX image of the sample, as shown in Figure~\ref{ch}(g), predicts the actual elemental composition of MoSeTe. The obtained atomic percentage ratios are consistent with the stoichiometry of MoSeTe. Furthermore, the TEM image shown in Figure~\ref{ch}(h) indicates randomly oriented particles with different shapes and sizes. In Figure~\ref{ch}(i) we show the magnified image in which the fringes are clearly observed and the $d-$spacing is found to be about 0.41~nm with the help of ImageJ software. The inter-layer spacing is found to be significantly larger to accommodate the sodium ion and providing the easy migration for battery applications. 

\begin{figure*}
     \centering
     \includegraphics[width=14cm,height= 10cm]{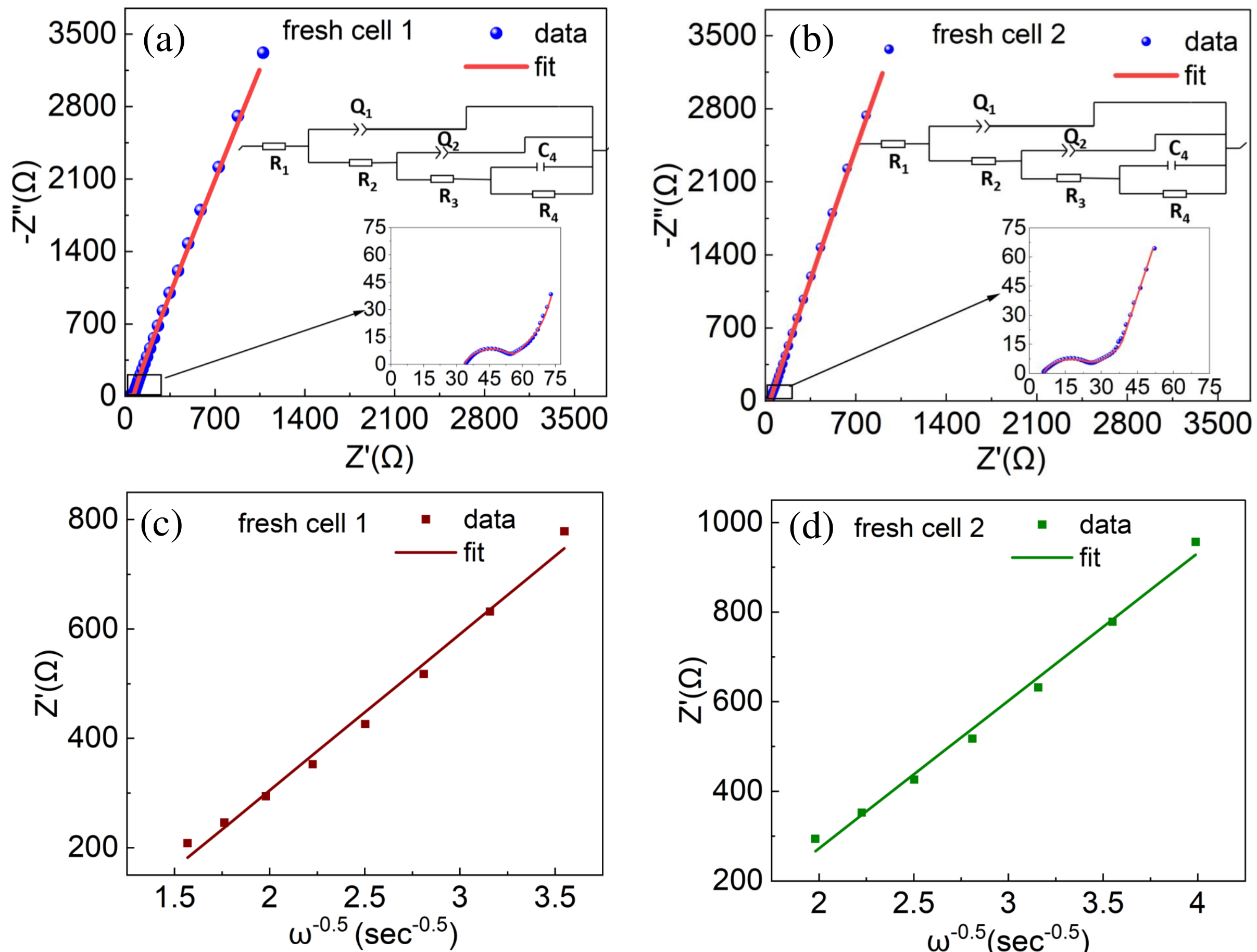}\\
	 \renewcommand{\figurename}{Figure}
	\caption {(a, b) The Nyquist plot, electrochemical impedance spectra of MoSeTe fresh cells \#1, and \#2. The insets show zoomed view and fitted circuit. The linear fit of Z' with  $\omega$$ ^{-0.5}$ in low frequency range for fresh cells \#1 and \#2 in (c) and (d), respectively.}
	 \label{EIS1}
 \end{figure*}

Therefore, to check the suitability of MoSeTe as an anode for sodium-ion batteries, electrochemical investigations were carried out. In order to evaluate the electrochemical kinetics of as prepared MoSeTe material, the EIS tests of fabricated coin cells were performed. In the case of fresh cell of MoSeTe as anode material, Na metal foil as counter/reference electrode and Cu foil as a current collector, the value of charge transfer resistance is found to be very low. Figures~\ref{EIS1}(a, b) consist of a semi-circle in high-frequency region and an inclined line in low-frequency region. The semi-circle represents the charge transfer resistance at the interface and the latter indicates the diffusion of sodium ions through the layered structure of bulk MoSeTe electrode \cite{r3}. 
 
   \begin{figure*}
  \centering   
	\includegraphics[width=14cm,height= 10cm]{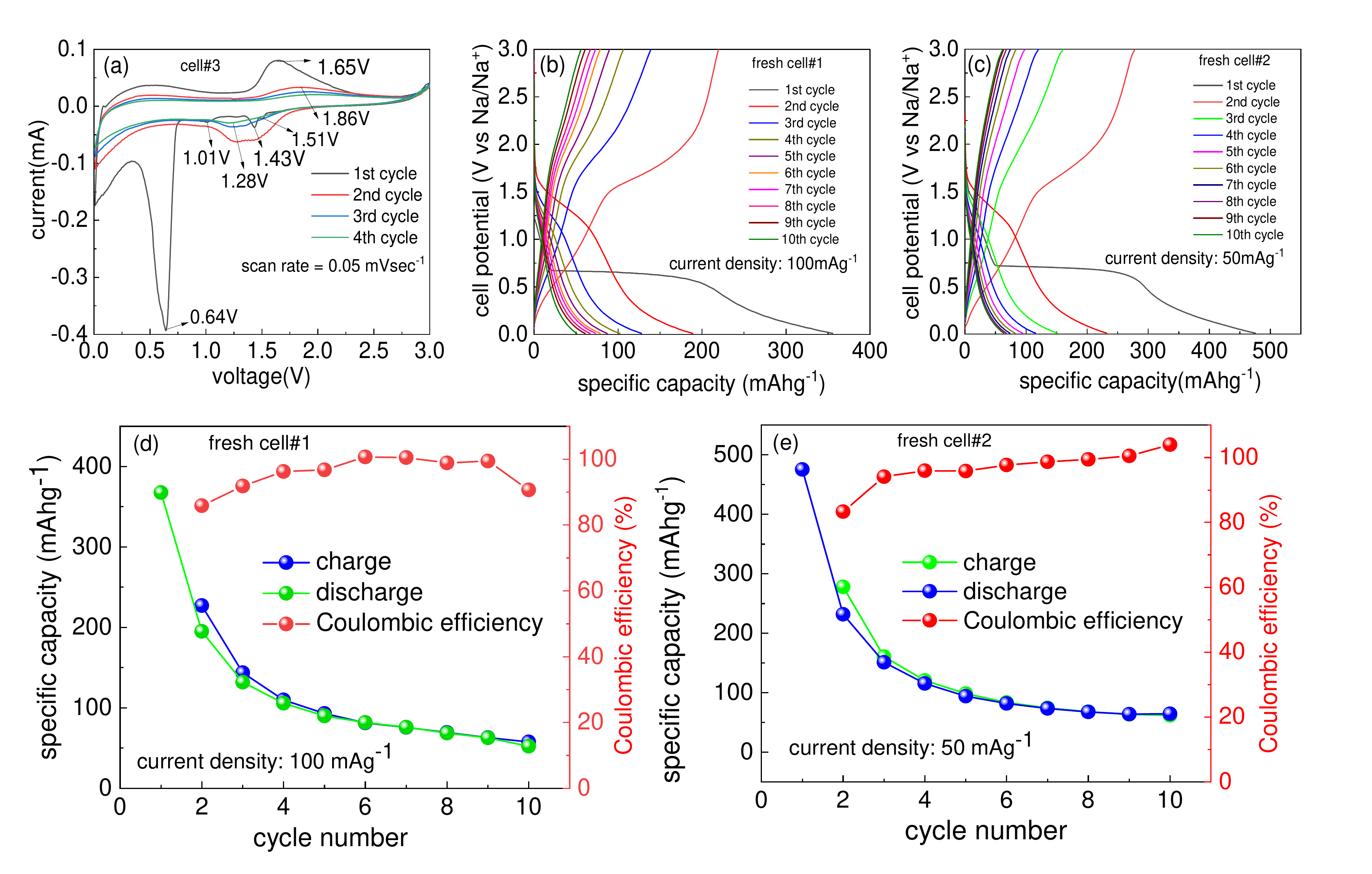} \\
	\renewcommand{\figurename}{Figure}
	\caption {The electrochemical measurements of MoSeTe: (a) the CV curves of fresh cell \#3 at a scan rate of 0.05~mVs$ ^{-1} $ for 4 cycles. The charge${-}$discharge profiles of fresh cells \#1 and \#2 for 10 cycles at current densities of (b) 100~mAg$^{-1}$ and (c) at 50~mAg$^{-1}$, respectively. The cycling performance and Coulombic efficiency of (d) fresh cell \#1 at current density of 100~mAg$^{-1}$ and (e) fresh cell \#2 at current  density of 50~mAg$^{-1}$.}
	 \label{CV}
 \end{figure*}

The Nyquist impedance plots between Z$_{re}$ (Z’) and --Z$_{im}$ (--Z”) at OCV shown in Figures~\ref{EIS1}(a, b) represent the MoSeTe fresh cells \#1 and \#2, respectively, in a frequency range of 100~kHz--10~mHz. To get more information about the electrochemical kinetics, the EIS curves were fitted where in Figure~\ref{EIS1}, the R$_1$ represents the solution resistance (R$_{s}$), R$_2$ represents the charge transfer resistance (R$_{ct}$) at the solid--electrolyte interface (SEI) and R$_3$ represents charge transfer resistance after the SEI layer formation \cite{r25}. Further, the R$_1$  (intersection point of high-frequency region semi-circle and Z$_{re}$ axis) is found to be 33.5~\ohm, and the R$_2$ provides the charge transfer resistance of 55.8~\ohm~ for the fresh cells \#1. Similarly, for fresh cell \#2, the value of solution resistance (R$_1$) is found to be 5.8~\ohm, while the value of charge transfer resistance R$_2$ is 29.8~\ohm, as shown in Figure~\ref{EIS1}. Moreover, the R$_4$ illustrates the charge transfer resistance at grain boundaries in the bulk electrode. The Q$_1$ and Q$_2$ are the circuit elements indicating the constant phase element owing to the imperfect capacitor behavior of the system. Whereas, the C$_4$ represents the behavior of the electrical double layer capacitor of the circuit in the bulk electrode. Further, to understand the diffusion kinetics, we have calculated the diffusion coefficient with the help of below equation \cite{r4}: 
\begin{equation}
D_{\mathrm{Na}^{+}}=R^{2} T^{2} / 2 n^{4} F^{4} \sigma_{\mathrm{w}}^{2} A^{2} C^{2}
\label{7} 
\end{equation}         
where $R$ is gas constant, $T$ is absolute temperature, $A$ is the area of the used electrode (1.76 cm$^{2}$), $F$ indicates Faraday’s constant (96500~C), $n$ ($n=$ 4) shows the number of electrons participating in electrochemical reaction, $C$ is bulk concentration of sodium ions (0.001~mole/cc) and $\sigma$ is slope of linear fit between real part of impedance(Z’) and $\omega$$^{-0.5}$ using equation (\ref{8}):
\begin{equation}
Z^{\prime}=R_{s}+R_{c t}+\sigma \omega^{-0.5}
\label{8} 
\end{equation}                                      
The plots between real part of impedance(Z’) and $\omega$$^{-0.5}$ are shown in Figures~\ref{EIS1}(c, d) for both the fresh cells \#1 and \#2, respectively. In equation~2, the R$_s$ and $\omega$ represent the ohmic resistance/solution resistance and angular frequency, respectively \cite{r5}. The diffusion coefficient of sodium ion in case of MoSeTe fresh cells \#1 are \#2 is found to be 2.2×10$^{-14}$ cm$^{2}$sec$^{-1}$  and 3.6×10$^{-14}$ cm$^{2}$sec$^{-1}$ having R$_{ct}$ values equal to 55.8~\ohm~ and 29.8~\ohm, respectively. The order of the diffusion coefficient depicts sluggish diffusion of sodium-ions at open circuit voltage measurements.   

 \begin{figure*}
     \centering
     \includegraphics[width=14cm,height= 10cm]{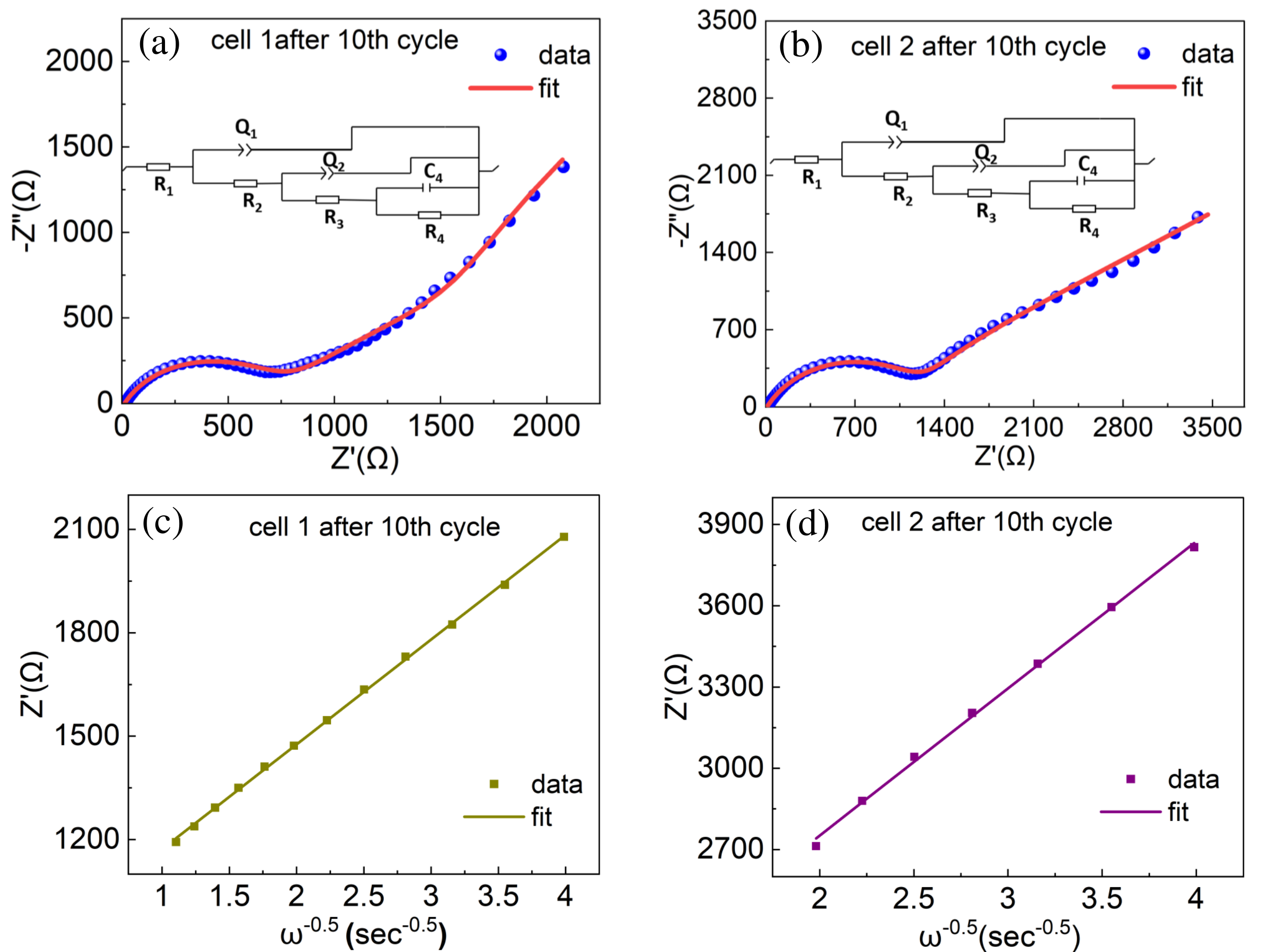}
	\renewcommand{\figurename}{Figure}
	\caption {The Nyquist plot, electrochemical impedance spectra of the MoSeTe cells (a) \#1 at 100 mAg$^{-1}$, and (b) \#2 at 50 mAg$^{-1}$ after 10th cycle of charging-discharging. The insets show the fitted circuit. (c, d) The linear fit of Z' with $\omega$$ ^{-0.5} $ in low frequency range.} 
	 \label{EIS2}
 \end{figure*}

Further, as prepared MoSeTe material was tested using cyclic voltammetry (CV) analysis where the measurements were performed in a voltage window of 0.01 to 3~V at a scan rate of 0.05~mVsec$^{-1}$. Figure~\ref{CV} shows the CV curves of fresh cell \#3 measured up to 4 cycles in which the reduction peaks at about 1.43~V and 1.51~V in first cathodic sweep can be ascribed to the formation of $Na_{x}$MoSeTe, as written in the  equation (\ref{9}):  
 \begin{equation}
xNa +MoSeTe   = Na_{x}MoSeTe
\label{9} 
\end{equation}
The evolution of reduction peak at 1.01~V is due to further insertion of sodium ion in combination with the formation of a layer of solid electrolyte interphase (SEI) \cite{r6}. The other reduction peak at 0.64~V is ascribed to the reduction process of $Na_{x}$MoSeTe into Mo metal and the formation of SEI films as per the conversion reaction mechanism written in equation (\ref{10}) \cite{r7, r9}:  
 \begin{equation}
Na_{x}MoSeTe {+} (4 - x) Na = Mo + Na_{4}SeTe \hspace{0.5mm} (x{<}2) 
\label{10} 
\end{equation}
In the anodic process, the peak at 1.65~V corresponds to the conversion of Mo to MoSeTe \cite{r10,r7}.  In the subsequent process, from cycle 2nd to cycle 4th, the peaks of the CV curves are slightly shifted from the 1st cycle due to some irreversible reactions and the SEI formation \cite{r11,r4}. The oxidation and reduction peaks from the 2nd to 4th cycle remain at the same initial position 1.86~V and 1.28~V, respectively, illustrating the good stability of the electrochemical process \cite{r12,r13}.

 \begin {table*}
\caption {The electrochemical performance of different TMD anode materials.} 
\label{tab:title} 
\begin{center}
\begin{tabular}{ | >{\centering\arraybackslash}m{2in} | >{\centering\arraybackslash}m{1in} | >{\centering\arraybackslash}m{2in} |}
 \hline
 \textbf{materials} & \textbf{Electrolyte} & \textbf{performence}\\  \hline
 MoS$_2$ (exfoliation) \cite{SU_AEM_15} &  1M NaClO$_4$ in EC/PC (1:1)& 386 mAh/g after 100 cycle at 0.04 A/g\\ 
 \hline
MoS$_2$  (liquid-phase exfoliation)\cite{BANG_API_14} &  1M NaClO$_4$ in PC/FEC & 161 mAh/g after 50 cycle  at 0.02 A/g\\ \hline
MoS$_2$  (CVD)\cite{COOK_AEM_16} &  1M NaClO$_4$ in PC/FEC & 520 mAh/g in 100 cycle at 1 A/g\\ 
\hline
MoS$_2$  (hydrothermal)\cite{GAO_NANO_17} & 1M NaClO$_4$ in EC/DMC (1:1) &370 mAh/g in 200 cycle at 2 A/g\\ 
\hline
 MoSe$_2$  (colloidal)\cite{WANG_JES_16} &  1M NaClO$_4$ in EC/DMC (1:1)& 345 mAh/g in 200 cycle at 42.2 mA/g\\ 
 \hline
 MoSe$_2$ \cite{ZHANG_JPS_15} &  1M NaClO$_4$ in EC/PC with 5\% FEC (1:1)& 450 mAh/g in 90 cycle at 200 mA/g \\ 
 \hline
 MoTe$_2$ \cite{PANDA_NE_19} &  1M NaClO$_4$ in EC/PC (1:1) with 3\% FEC&270 mAh/g in 200 cycle at 1 A/g \\ 
 \hline
VS$_2$  (solvothermal)\cite{ZHOU_AM_17} &  1M NaClO$_4$ in DEC/EC (1:1) with 6\% FEC& 700 mAh/g in 100 cycle at 0.1 A/g\\ \hline
VS$_2$  (hydrothermal)\cite{WANG_JMC_19} &  1M NaSO$_3$CF$_3$ in DGM & 565 mAh/g in 1200 cycle at 2.0 A/g\\ \hline
NbSe$_2$ (solid-state sintering)\cite{XU_IONIC_19} &  1M NaClO$_4$ in DEC/EC (1:1)& 98.1 mAh/g in 100 cycle at 0.1 A/g\\ \hline
NbS$_2$  (chemical exfoliation)\cite{OU_JPS_16} &  1M NaPF$_6$ in EC/DMC (1:1)& 157 mAh/g in 100 cycle at 0.5 A/g\\ \hline
WTe$_2$  (hydrothermal)\cite{HONG_EF_18} &  1M NaClO$_4$ in DEC/EC (1:1) with 5\% FEC& 221 mAh/g in 100 cycle at 0.1 A/g\\ \hline
WTe$_2$  (CVD)\cite{HONG_EF_18} &  1M NaClO$_4$ in DEC/EC (1:1) with 5\% FEC & 260 mAh/g in 40 cycle at 0.1 A/g\\ 
\hline
\end{tabular}
\end{center}
\end {table*}

In order to understand the sodiaton/de-sodiation mechanism in the MoSeTe anode, the galvanostatic charge-discharge (GCD) measurements were carried out within the voltage range 0.01 to 3~V at the current densities of 100~mAg$^{-1}$ (cell \#1) and 50~mAg$^{-1}$ (cell \#2). The voltage plateau appears below 1~V is associated with the sodiation of MoSeTe and the SEI layer formation \cite{r4}. In case of the cell \#1, the value of first discharge capacity is 355 mAhg$^{-1}$ and second charge, discharge capacities are 219, 188 mAhg$^{-1}$, respectively at a current density of 100~mAg$^{-1}$, as shown in Figure~\ref{CV}(b). Also, for the cell \#2, the value of first discharge capacity is 475 mAhg$^{-1}$ and second charge, discharge capacities are 278, 231 mAhg$^{-1}$, respectively at a current density of 50~mAg$^{-1}$, as shown in Figure~\ref{CV}(c).The difference between first discharge and charge  capacity is due to the formation of SEI layer \cite{r14, r15, r16}. In Figures~\ref{CV}(d, e), we show the cyclic measurements and Coulombic efficiency for both the cells \#1 and \#2, respectively. We notice that the capacity is degrading faster where only about 30\% retention is observed after 10 cycles, but it is important to mention that the Coulombic  efficiency found to be  $\approx$100\%, which motivate for further modification of this anode material to improve the stability. The significant capacity fading is an indication of surface instability during sodiation/de-sodiation in the host structure, which is also consistent with the EIS results after the 10$^{th}$ cycle, as shown in Figures~\ref{EIS2}(a, b) and discussed below. Here, the value of first discharge capacity in case of 50 mAg$^{-1}$ is greater than that of 100 mAg$^{-1}$ owing to large exposure time for sodium-ions insertion/de-insertion process in the host material at lower current density. The continuous process of sodium-ion insertion-extraction deep within the host material is causing irreversible distortion, because of the large in-plane stiffness present in the material. The stiffness is defined as force per unit deformation, where the externally applied potential forces the sodium ions to intercalate into the host structure, which causes large stiffness in the host material retarding the modification in the structure for accommodating sodium ions. This forceful intercalation and resistance to structure modification is breaking the structure of MoSeTe irreversibly after certain charge-discharge cycles. This phenomenon can be interpreted from the constant capacity fading of the material during GCD with increasing the number of cycles. The substantial capacity fading observed was most likely caused by the loss of active sodium-ions generated by the traps within the anode material. 

Moreover, to evaluate the diffusion kinetics of MoSeTe after 10th cycle of charging-discharging, the EIS tests were performed again. The value of charge transfer resistance is found to be very high as compared to that of the fresh cells. Figures~\ref{EIS2}(a, b) show the Nyquist impedance plot between Z$_{re}$ (Z’) and --Z$_{im}$ (--Z”) in a frequency range of 100~kHz--10~mHz, which attribute the electrochemical kinetics of cell \#1 and cell \#2, respectively after 10th cycle of charging-discharging. In addition, the EIS curve fitting was done to collect more information about the effect of cycling on the sodium-ion insertion/de-insertion kinetics. As described earlier in Figures~\ref{EIS1}(a--d) all symbols have their own meaning. Further, the R$_1$ is found to be 11.9~\ohm, whereas the R$_2$ provides the charge transfer resistance of 771.6~\ohm~ for the cell \#1 after 10th cycle at a current density of 100 mAg$^{-1}$ and for cell \#2 after 10th cycle at 50 mAg$^{-1}$, the value of solution resistance is found to be 6.2~\ohm, and the R$_2$ is 1410~\ohm. After 10 cycles, the charge-transfer resistance of both cell \#1 and \#2 increases dramatically. This finding shows that during charge/discharge processes, the electrolyte continues to deteriorate and uniformly deposit on the anode's surface, increasing the anode's electrochemical polarization and resulting in diminishing battery capacity \cite{r17}. The EIS results revealed that the R$_{ct}$ of cell \#2 at 50~mAg$^{-1}$ was significantly higher than that of cell \#1 at 100~mAg$^{-1}$. This suggested that a higher current density would limit the available charge-discharge specific capacity, slowing the deterioration process. Hence, the higher current densities could mitigate the degradation because of the lower capacity utilization \cite{r18}. The R$_{ct}$ value of reported MoTe$_{2}$ \cite{ChoNS17} and MoSe$_{2}$ \cite{WANG_JES_16} before cycling is much higher than MoSeTe (present work), whereas the R$_{ct}$ after cycling for these materials are lower than that of MoSeTe. This feature indicates fast electron kinetics in MoSeTe at OCV, but there is surface instability due to high over-potential after cycling and therefore faster degradation of capacity. In addition, the diffusion coefficient is also calculated with the help of the equations (\ref{7}) and (\ref{8}) where the values are found to be 1.9×10$^{-14}$ cm$^{2}$sec$^{-1}$ and 0.3×10$^{-14}$ cm$^{2}$sec$^{-1}$ for the cells \#1 and \#2, respectively, after 10th cycle. This again indicates a sluggish diffusion of sodium-ions in the host structure. In Table~1, we compare the electrochemical performance of different TMD anode materials reported in literature.   
 
 \section*{Conclusions}
 
In a half-cell configuration, the MoSeTe as anode material exhibits significant initial specific discharge capacities of 475 and 355 mAhg$^{-1}$ at current densities of 50 and 100 mAg$^{-1}$, respectively. The Coulombic efficiency is found to be nearly 100\%; however, the capacity in 2nd cycle decreased to $\approx$200~mAhg$^{-1}$. The cyclic voltammetry (CV) study reveals the reversibility of the material after 1st cycle resulting in no change in the initial peak positions. The electrochemical impedance spectroscopy measurements affirm small charge transfer resistance at OCV for  the fresh cells, which provides a more conducting nature and easy diffusion of Na$^{+}$ between the electrode and electrolyte. However, the significant capacity fading after cycling needs to be improved by making MoSeTe@rGO composite and enhance its electrochemical performance by reducing overall resistance. 
 
 \section*{Acknowledgments }
 
  We acknowledge the financial support from the DST-IIT Delhi Energy Storage Platform on Batteries (DST/TMD/MECSP/2K17/07) and SERB-DST through a core research grant (CRG/2020/003436). We thank IIT Delhi for providing research facilities, and MHRD and UGC for the fellowship.
  


\end{document}